\documentstyle[preprint,tighten,pra,aps,latexsym,amssymb]{revtex}

\newcommand{\dd}{{\rm d}}
\newcommand{\ee}{{\rm e}}

\newcommand{\rif}[1]{(\ref{#1})}

\newcommand{\eq}{\begin{equation}}
\newcommand{\feq}{\end{equation}}
\newcommand{\eqn}{\begin{eqnarray}}
\newcommand{\feqn}{\end{eqnarray}}
\newcommand{\arr}{\begin{eqnarray*}}
\newcommand{\farr}{\end{eqnarray*}}

\newcommand{\lp}{\left(}
\newcommand{\rp}{\right)}

\newcommand{\R}{{\mathbb R}}

\begin{document}
\tightenlines
\draft

\def\al{\alpha}
\def\be{\beta}
\def\ga{\gamma}
\def\de{\delta}
\def\ep{\varepsilon}
\def\ze{\zeta}
\def\io{\iota}
\def\ka{\kappa}
\def\la{\lambda}
\def\roh{\varrho}
\def\si{\sigma}
\def\om{\omega}
\def\ph{\varphi}
\def\th{\theta}
\def\te{\vartheta}
\def\up{\upsilon}
\def\Ga{\Gamma}
\def\De{\Delta}
\def\La{\Lambda}
\def\Si{\Sigma}
\def\Om{\Omega}
\def\Te{\Theta}
\def\Th{\Theta}
\def\Up{\Upsilon}

\preprint{UTF 420}

\title{Supersymmetry of Anti-de Sitter Black Holes}

\author{Marco M.~Caldarelli\footnote{email: caldarel@science.unitn.it}
and Dietmar Klemm\footnote{email: klemm@science.unitn.it}\\ \vspace*{0.5cm}}

\address{Universit\`a  degli Studi di Trento,\\
Dipartimento di Fisica,\\
Via Sommarive, 14\\
38050 Povo (TN)\\
Italia\\
\vspace*{0.5cm}      
and\\ Istituto Nazionale di Fisica Nucleare,\\
Gruppo Collegato di Trento,\\ Italia}

\maketitle
\begin{abstract}
We examine supersymmetry of four-dimensional asymptotically
Anti-de Sitter (AdS) dyonic black holes in the context of gauged $N=2$
supergravity. Our calculations concentrate on black holes with unusual
topology and their rotating generalizations, but we also reconsider
the spherical rotating dyonic Kerr-Newman-AdS black hole, whose supersymmetry
properties have previously been investigated by Kosteleck\'{y} and Perry
within another approach. We find that in the case of spherical, toroidal
or cylindrical event horizon topology, the black holes must rotate in
order to preserve some supersymmetry; the non-rotating
supersymmetric configurations representing naked singularities.
However, we show that this is no more true for black holes whose
event horizons are Riemann surfaces of genus $g>1$, where we find a
nonrotating extremal solitonic black hole carrying magnetic charge
and permitting one Killing spinor. For the nonrotating supersymmetric
configurations of various topologies, all Killing spinors are
explicitely constructed.
\end{abstract}


\maketitle

\section{Introduction}
Black holes in Anti-de Sitter space represent a subject of current
interest, which, on the one hand, is based on Maldacena's conjecture of
AdS/Conformal field theory-correspondence \cite{malda}, and, on the other
hand, on the fact that AdS space admits black holes with unusual
topology \cite{amin,mann,lemos,zhang,huang,vanzo}.
These so-called topological black
holes have some intriguing properties \cite{kv}.
Among them are the uncommon behaviour of
the luminosity, another version of the information loss paradoxon
due to the boundary conditions necessary in an asymptotically AdS space,
and a mass spectrum that seems to be difficult to reconcile with string theory.
In fact, the argument of Horowitz-Polchinski \cite{horpol}, which
describes the transition from a highly excited string state to a black hole,
and provides a microscopic interpretation of black hole entropy,
does not seem to work for topological black holes, at least not in a naive
manner \cite{kv}.
The reason for this is that the mass levels of a
string in AdS space are $M \approx \ell^{-1}n$ for large quantum numbers
$n$ \cite{larsen}, $\ell$ being related to the
cosmological constant via $\Lambda = -3/\ell^{2}$.
This yields a black hole entropy proportional to $n$, whereas the string
entropy goes with $\sqrt{n}$ (c.~f.~\cite{kv} for details).
So it would seem that a string had not enough degrees of freedom to
account for black hole entropy. However,
the correspondence principle of Horowitz-Polchinski, which has turned
out to be very successful up to now,
can not be rejected on the above, rather naive arguments,
for example there are other mass spectrum regimes for a string in
AdS (for $\ell/l_s \gg 1$, where $l_s$ is the string length, the mass
spectrum is like in flat space), or there could be a transition from
a configuration of D-branes to a black hole. In short, it remains to
see how exactly the argument of Horowitz-Polchinski works for topological
black holes, and it seems quite
challenging to try to give a microscopic
description of the entropy of these objects within string theory,
e.~g.~by using D-brane technology. A first step in this enterprise is to
find supersymmetric configurations, as for BPS states we know that the
degeneracy at weak string coupling constant $g_s$ does not change if one
increases $g_s$. A natural candidate to address the issue of supersymmetry
of topological AdS black holes is $N=2$ gauged supergravity
\cite{dasfreed,fradvas}. In this theory,
the rigid $\mathrm{SO(2)}$ symmetry, rotating the 2 independent Majorana
supersymmetries present in the ungauged theory, is made local. This
requires a negative cosmological constant.\\
Supersymmetry of AdS black holes with spherical event horizons have been
studied before in the literature
\footnote{For BPS black holes in five-dimensional $N=2$ gauged supergravity
coupled to vector supermultiplets
cf.~\cite{behrndt}.}.
Romans \cite{romans} showed that
the Reissner-Nordstr\"om-AdS black hole is supersymmetric in two cases.
The first one appears for $q_m=0$ and $q_e^2=m^2$; $q_m$, $q_e$ and $m$
being the magnetic charge, electric charge and mass parameter,
respectively. The second one emerges for $m=0$ and $q_m=\pm \ell/2$
(we recall that $\Lambda=-3/\ell^2$ is the cosmological constant). However,
all these supersymmetric configurations represent naked singularities.
The situation is similar to the asymptotically flat Kerr-Newman
black hole, which reaches the extreme limit $m^2=a^2 + q_e^2$
($a$ denoting the rotation parameter) before the
supersymmetry condition $m^2=q_e^2$ is satisfied \cite{tod}.\\
As far as the spherical Kerr-Newman-AdS solution is concerned,
it was shown by Kosteleck\'{y} and Perry \cite{perry} that only for
nonvanishing rotation parameter $a$ it is possible to obtain
supersymmetric extremal black holes. This means that in an AdS
background, rotating black holes are the analogue of the asymptotically
flat extreme Reissner-Nordstr\"om solution.
This result was obtained by considering the Bogomol'nyi bound arising from
the supersymmetry algebra $\mathrm{osp(4|2)}$ of gauged $N=2$ supergravity.
It is given by $m = \sqrt{q_e^2 + q_m^2}(1 \pm a/\ell)$. In the present
paper we will show that in this equation the magnetic charge is required
to be zero, making thus the Kosteleck\'{y}-Perry result more precise.
Besides, we will see that there exists also a supersymmetric configuration
with vanishing mass parameter, but nonvanishing magnetic charge. This
represents a naked singularity, and it was not obtained in \cite{perry}.\\
The main purpose of our paper, however, is to study supersymmetry of
black holes with unusual topology, which has not been considered in the
literature before. The rest of this article is organized as follows:\\
In section \ref{topbh} we give a short introduction into the geometry of
four-dimensional topological black holes and classify all known solutions.\\
In section \ref{sugra} gauged $N=2$, $d=4$ supergravity is briefly reviewed.\\
In section \ref{nonrot} we investigate supersymmetry of nonrotating
black holes whose event horizons are Riemann surfaces of genus $g \ge 1$.
The Killing spinors in the various cases are explicitely constructed.\\
In section \ref{rot} we generalize our results to rotating black holes
of various topologies.\\
Finally, our results are summarized and discussed in \ref{disc}.

\section{Anti-de~Sitter Black Holes}\label{topbh}

In this section, we shall review four-dimensional
asymptotically AdS black holes,
which are solutions of the Einstein-Maxwell equations with negative
cosmological constant. The cosmological constant is sufficient to avoid a
few classic theorems forbidding nonspherical black holes \cite{hawking,fsw}. 
As a result,
beyond the well-known Kerr-Newman-AdS black hole, there is a huge variety
of black holes with unusual topology. We shall first show how topological
black holes arise in AdS space in the simple nonrotating case, then we shall
consider the rotating generalizations.
Throughout the discussion, particular attention
will be paid on the extreme black holes. Although not explicitly stated in
the following, all the metrics we shall discuss are also solutions of
$N=2$ gauged supergravity, as will become clear in section \ref{sugra}.

\subsection{Nonrotating AdS Black Holes}

We start from the class of metrics
\eq
\dd s^2=-V(r)\dd t^2+V^{-1}(r)\dd r^2+r^2\dd\sigma^2,
\label{NonRot}
\feq
where $\dd\sigma^2$ is the metric of a two-manifold $\cal S$. The
Einstein-Maxwell equations with negative cosmological constant $\Lambda =
-3\ell^{-2}$ require
$\cal S$ to be a surface of constant curvature $\kappa$, and
\eq
V(r)=\kappa-{2\eta\over r}+{z^2\over r^2}+{r^2\over \ell^2},
\qquad z^2=q_e^2+q_m^2, \label{lapsegen}
\feq
where $q_e$ and $q_m$ denote the electric and magnetic charge parameters
respectively.
It is useful to define
\eq
\eta_0(z)=\frac{\ell}{3\sqrt{6}}\lp\sqrt{\kappa^2+12z^2\ell^{-2}} +
2\kappa\rp\lp\sqrt{\kappa^2+12z^2\ell^{-2}}-\kappa\rp^{1/2}.
\feq
According to the sign of the curvature of $\cal S$ we obtain three cases:
\begin{enumerate}
\item $\kappa=1$, $\dd\sigma^2=\dd\theta^2+\sin^2\theta\ \dd\phi^2$
  corresponds to the charged spherical AdS black hole 
  if $\eta\geq\eta_0(z)$, to
  AdS space if $\eta=z=0$ and otherwise to a naked singularity, 
\item $\kappa=0$, $\cal S$ flat, describes a flat charged black membrane in
  AdS when $\eta\geq\eta_0(z)>0$. A genuine black hole
  should have a compact orientable event horizon, hence we should take some
  quotient of $\cal S$. As a result, $\dd\sigma^2=\dd x^2+2{\rm Re }\tau\dd
  x\dd y+|\tau|^2\dd y^2$, $x,y\in[0,1]$ with $0$ and $1$ identified, 
  and $\cal S$ is a torus with complex Teichm\"uller parameter $\tau$, and
  the metric \rif{NonRot} describes a charged toroidal black hole. For
  $\eta<\eta_0(z)$, as well as for $\eta=z=0$, the spacetime has a naked
  singularity,
\item $\kappa=-1$, $\dd\sigma^2=\dd\theta^2+\sinh^2\theta\ \dd\phi^2$,
  $\cal S$ is the hyperbolic plane $H^2$,
  and when $\eta\geq\eta_0(z)$ we are again
  dealing with a
  charged black membrane in AdS. As is well known, $H^2$ is the
  universal covering space for all Riemann surfaces of genus
  $g>1$. Quotienting $\cal S$ with a suitable discrete subgroup of
  its isometry group $\mathrm{SO}(2,1)$,
  the metric \rif{NonRot} will describe charged
  higher genus black holes. For $\eta<\eta_0(z)$ the spacetime has a naked
  singularity.
\end{enumerate}
The electromagnetic potential one-form is given by
\eq
A=-\frac{q_e}r\dd t+q_m\cos\theta\dd\phi,\quad 
A=-\frac{q_e}r\dd t+q_m|{\rm Im }\tau|x\dd y,\quad 
A=-\frac{q_e}r\dd t+q_m\cosh\theta\dd\phi,
\feq
for the sphere, torus and higher genus case
respectively.\\
The causal structure of these black holes has been studied in \cite{BLP},
and we refer to that paper for the Penrose diagrams. In the three cases,
for $\eta>\eta_0(z)>0$
the black hole has an outer event horizon and an internal Cauchy horizon; the
singularity is timelike, in analogy with the Reissner-Nordstr\"om black hole.
For $z=0$, $\eta>0$, the black hole has a simple event horizon hiding a
spacelike singularity, while for $\eta=\eta_0(z)>0$ the lapse function has a
double root, and the black hole is extreme. In addition, in the higher
genus $\kappa=-1$ case, there is an uncharged nonrotating extreme black
hole for $(\eta=-\ell/3\sqrt 3, z=0)$, black holes with inner and outer
horizons for $(-\ell/3\sqrt 3<\eta<0,z=0)$ and a locally AdS black hole with a
single horizon for $\eta=z=0$.\\
The computation of the mass of these black holes involves some subtleties,
as a proper choice of the reference background has to be done \cite{vanzo};
in the spherical and toroidal cases the appropriate background is that
obtained by putting $\eta$ and the charges to zero, while for
higher genus black holes one has to take the uncharged extreme black hole
with mass parameter $\eta_0=-\ell/3\sqrt{3}$ as background, to have the
Arnowitt-Deser-Misner (ADM) mass as a positive, concave function of the
black hole's temperature as defined by its surface gravity.
Taking this into account, one obtains
\eq
M=\eta,\qquad M={\eta|{\rm Im }\tau|\over 4\pi},\qquad M=(\eta-\eta_0)(g-1),
\label{mass}
\feq
for the mass of the spherical, toroidal, and genus $g>1$ black holes
respectively.\\
The total electric charge in the various cases is
\eq
Q_e=q_e,\qquad Q_e=q_e|{\rm Im }\tau|,\qquad Q_e=q_e(g-1),
\feq
and the magnetic charge
\eq
Q_m=q_m,\qquad Q_m=q_m|{\rm Im }\tau|,\qquad Q_m=q_m(g-1). \label{magncharge}
\feq
\\
For $\eta=z=0$, the metric \rif{NonRot} is locally AdS; for $\kappa=1$ we
obtain AdS space, for $\kappa=0$ a quotient of AdS with a naked
singularity, and finally, for $\kappa=-1$ we obtain a quotient of AdS space
with a black hole interpretation \cite{amin}, a four-dimensional analogue
of the Ba\~nados-Teitelboim-Zanelli (BTZ) black hole \cite{BTZ}.\\
The properties of these black holes have been extensively investigated in
recent times. They can form by gravitational collapse \cite{mann},
they emit Hawking radiation \cite{kv}, and a consistent thermodynamics can
be formulated for them: they respect the zeroth and first law, and obey the
entropy-area law \cite{vanzo,BLP}.

\subsection{Kerr-Newman-AdS Black Hole}

This is the usual charged rotating black hole in AdS. Its horizon is
diffeomorphic to a sphere, and its metric, which is axisymmetric, reads in
Boyer-Lindquist-type coordinates
\eq
\dd s^2 = -{\Delta_r\over\Xi^2\rho^2}
\left[\dd t-a\sin^2\theta\ \dd\phi\right]^2
+{\rho^2\over\Delta_r}\dd r^2+{\rho^2\over\Delta_\theta}\dd\theta^2
+{\Delta_\theta\sin^2\theta\over\Xi^2\rho^2}
\left[a\dd t-(r^2+a^2)\dd\phi\right]^2,
\label{KNAdS}\feq
where
\eq 
\rho^2=r^2+a^2\cos^2\theta, \qquad \Xi=1-{a^2\over \ell^2},
\feq
\eq
\Delta_r=(r^2+a^2)\lp 1+{r^2\over \ell^2}\rp-2mr+z^2 , \qquad 
\Delta_\theta=1-{a^2\over \ell^2}\cos^2\theta ,
\feq
$a$ is the rotational parameter and $z$ is defined by $z^2=q_e^2+q_m^2$,
with $q_e$ and $q_m$ the electric and magnetic charge parameter respectively. 
This metric solves the Einstein-Maxwell field equations with an
electromagnetic vector potential given by
\begin{eqnarray}
A&=&-{q_er\over\Xi\rho^2}\left[\dd t-a\sin^2\theta\dd\phi\right]
-{q_m\cos\theta\over\Xi\rho^2}\left[a\dd t-(r^2+a^2)\dd\phi\right]\nonumber\\
&=&-{q_er\over\rho\sqrt{\Delta_r}}\ {\rm e}^0
-{q_m\cos\theta\over\rho\sqrt{\Delta_\theta}\sin\theta}\ {\rm e}^3,
\end{eqnarray}
where $\ee^a$ is the vierbein (see Appendix \ref{appKNAdS}). 
The associated field strength tensor is
\begin{eqnarray}
F&=&-\frac 1{\rho^4}\left[ q_e(r^2-a^2\cos^2\theta)+2q_mra\cos\theta\right]
\ \ee^0\wedge\ee^1\nonumber\\
&&+\frac 1{\rho^4}\left[q_m(r^2-a^2\cos^2\theta)-2q_era\cos\theta\right]
\ \ee^2\wedge\ee^3.\label{fieldKNAdS}
\end{eqnarray}
Let us define the critical mass parameter $m_{extr}$,
\eqn
m_{extr}(a,z)&=&{\ell\over3\sqrt6}\lp\sqrt{\lp1+{a^2\over \ell^2}\rp^2+{12\over
    \ell^2}(a^2+z^2)}+{2a^2\over \ell^2}+2\rp\nonumber\\
&&\times\lp\sqrt{\lp1+{a^2\over \ell^2}\rp^2+{12\over
    \ell^2}(a^2+z^2)}-{a^2\over \ell^2}-1\rp^{1\over 2}.
\label{extrKNAdS}\feqn
A study of the positive zeroes of the lapse function $\Delta_r$ shows
that the metric \rif{KNAdS} describes
a naked singularity for $m<m_{extr}$ and a
black hole with an outer event horizon and an inner Cauchy horizon for
$m>m_{extr}$. Finally, for $m=m_{extr}$, the lapse function has a double root
and \rif{KNAdS} represents an extremal black hole.\\
Observing that $\partial_t$ and $\partial_\phi$ are Killing vectors, one can
use Komar integrals to define mass and angular momentum
of the Kerr-Newman-AdS black hole computed with respect to AdS
space. For the results, as well as for the electric and magnetic
charges, we refer to \cite{perry}.

\subsection{Rotating Generalization of the Charged
$g>1$ Topological Black Holes}

A rotating generalization of the topological black holes with genus
$g>1$ has been
obtained from the Kerr-AdS black hole by an analytical continuation \cite{KMV}.
Proceeding analogously from \rif{KNAdS} (leaving in addition the charge $z$
unaffected by the analytical continuation), we easily obtain a charged
generalization of the rotating topological black hole. The metric is given
by
\eq
\dd s^2 = -{\Delta_r\over\Xi^2\rho^2}
\left[\dd t+a\sinh^2\theta\ \dd\phi\right]^2
+{\rho^2\over\Delta_r}\dd r^2+{\rho^2\over\Delta_\theta}\dd\theta^2
+{\Delta_\theta\sinh^2\theta\over\Xi^2\rho^2}
\left[a\dd t-(r^2+a^2)\dd\phi\right]^2 ,
\label{HG}
\feq
where now
\eq 
\rho^2=r^2+a^2\cosh^2\theta , \qquad \Xi=1+{a^2\over \ell^2} ,
\feq
\eq
\Delta_r=(r^2+a^2)\lp-1+{r^2\over \ell^2}\rp-2\eta r+z^2 , \qquad 
\Delta_\theta=1+{a^2\over \ell^2}\cosh^2\theta .
\feq
Again, $a$ is the rotational parameter and $z$ is defined by $z^2=q_e^2+q_m^2$.
The metric \rif{HG} is of Petrov type D, obtained as a
special case of the most general kown type D solution found by
Plebanski and Demianski \cite{plebdem}. 
\rif{HG} solves the Einstein-Maxwell field equations with an
electromagnetic vector potential given by
\begin{eqnarray}
A&=&-{q_er\over\Xi\rho^2}\left[\dd t+a\sinh^2\theta\dd\phi\right]
-{q_m\cosh\theta\over\Xi\rho^2}\left[a\dd t-(r^2+a^2)\dd\phi\right]\nonumber\\
&=&-{q_er\over\rho\sqrt{\Delta_r}}\ {\rm e}^0
-{q_m\cosh\theta\over\rho\sqrt{\Delta_\theta}\sinh\theta}\ {\rm e}^3,
\end{eqnarray}
and the associated field strength tensor is
\begin{eqnarray}
F&=&-\frac 1{\rho^4}\left[ q_e(r^2-a^2\cosh^2\theta)+2q_mra\cosh\theta\right]
\ \ee^0\wedge\ee^1\nonumber\\
&&-\frac 1{\rho^4}\left[ q_m(r^2-a^2\cosh^2\theta)-2q_era\cosh\theta\right]
\ \ee^2\wedge\ee^3.\label{fieldHG}
\end{eqnarray}
\\
(For the vierbein $\mathrm{e}^a$ cf.~Appendix \ref{appHG}).
The coordinates $t$ and $r$ range over $\R$, while $\theta\in\R^+$ and
$\phi\in[0,2\pi]$ (with endpoints identified) parametrize the
sections of constant $(t,r)$
in polar coordinates. Hence, our solution describes a
charged rotating black membrane in AdS space.\\
The causal structure of these objects is very complicated and we are not
interested in a complete analysis (see \cite{KMV} for the study of the
causal structure in the uncharged case).\\
Also the thermodynamic behaviour of these black membranes
remains an open question,
which will be discussed elsewhere.

\subsection{Charged Rotating Cylindrical Black Hole}

A rotating generalization of the toroidal black hole cannot be found by
analytical continuation or by similar tricks, but it can be obtained from the
general Petrov type D solution, by an appropriate choice of parameters
\cite{KMV}. Allowing in addition nonvanishing electromagnetic charges, we 
obtain a rotating generalization of the charged toroidal black hole. The
metric is given by
\eq
\dd s^2 = -{\Delta_r\over\rho^2}
\left[\dd t+aP^2\dd\phi\right]^2
+{\rho^2\over\Delta_r}\dd r^2+{\rho^2\over\Delta_P}\dd P^2
+{\Delta_P\over\rho^2}
\left[a\dd t-r^2\dd\phi\right]^2,\label{rottorus}
\feq
where
\eq 
\rho^2=r^2+a^2 P^2,
\feq
\eq
\Delta_r=a^2+z^2-2\eta r+{r^4\over \ell^2} , \qquad 
\Delta_P=1+{a^2\over \ell^2}P^4.
\feq
As usual, $a$ is the rotational parameter and $z$ is defined by
$z^2=q_e^2+q_m^2$.
This metric is a solution of the
Einstein-Maxwell field equations with an electromagnetic vector potential
given by
\begin{eqnarray}
A&=&-{q_er\over\rho^2}\left[\dd t+aP^2\dd\phi\right]
-{q_mP\over\rho^2}\left[a\dd t-r^2\dd\phi\right]\nonumber\\
&=&-{q_er\over\rho\sqrt{\Delta_r}}\ {\rm e}^0
-{q_mP\over\rho\sqrt{\Delta_P}}\ {\rm e}^3,
\end{eqnarray}
and a field strength tensor
\begin{eqnarray}
F&=&-\frac 1{\rho^4}\left[ q_e(r^2-a^2P^2)+2q_mraP\right]
\ \ee^0\wedge\ee^1\nonumber\\
&&-\frac 1{\rho^4}\left[ q_m(r^2-a^2P^2)-2q_eraP\right]
\ \ee^2\wedge\ee^3,\label{fieldrottorus}
\end{eqnarray}
where the vierbein $\ee^a$ is given in appendix \ref{approttorus}.\\
For a given rotational parameter $a$ and charge parameter $z$ we define the
critical mass parameter
\eq
\eta_{extr}(a,z)={2\over3^{3/4}\ell^{1/2}}\lp a^2+z^2\rp^{3/4}.
\feq
If $\eta<\eta_{extr}$, $\Delta_r$ has no positive root and the metric
\rif{rottorus} describes a naked singularity.
For $\eta=\eta_{extr}$, there is a
double root in $\Delta_r$ and we obtain an extremal black hole
with a timelike singularity.
Finally, for $\eta>\eta_{extr}$ and $a^2+z^2\neq 0$,
$\Delta_r$ has two positive simple roots, and the metric (\ref{rottorus})
describes a black hole with an event horizon and an inner Cauchy horizon.
The Penrose diagrams can be
found in \cite{KMV}, substituting $a^2$ with $a^2+z^2$.\\
Now, if an horizon is present, it is not compact. The ranges of the
coordinates are $t,P\in\R$, $r\in\R^+$ and $\phi\in[0,2\pi]$ with the
extrema identified. Hence the black hole has a cylindrical event
horizon. For $a=0$, we can naturally compactify also the coordinate $P$
($\partial_P$ becomes a Killing vector in this case), whereupon
(\ref{rottorus}) reduces to the static toroidal black hole
spacetime (\ref{NonRot}) with $\kappa=0$.

\section{$N=2$ Gauged Supergravity} \label{sugra}
The gauged version of $N=2$ supergravity was found by Das and Freedman
\cite{dasfreed} and by Fradkin and Vasiliev \cite{fradvas}. In this theory,
the rigid $\mathrm{SO(2)}$ symmetry rotating the two independent
Majorana supersymmetries present in the ungauged theory, is made local
by introduction of a minimal gauge coupling between the photons and
the gravitini. Local supersymmetry then requires a negative cosmological
constant and a gravitini mass term.
The theory has four bosonic and four fermionic degrees of freedom;
it describes a graviton $e_m^a$, two real gravitini $\psi_m^i$ $(i=1,2)$,
and a Maxwell gauge field $A_m$. As we said, the latter is minimally
coupled to the gravitini, with coupling constant $\ell^{-1}$. If we
combine the two $\psi_m^i$ to a single complex spinor $\psi_m = \psi_m^1 +
i\psi_m^2$, the Lagrangian reads (cf.~also \cite{romans})
\footnote{Throughout this paper, the notation is as follows:
$a,b,\ldots$ refer to $d=4$ tangent space indices, and $m,n,\ldots$ refer to
$d=4$ world indices. The signature is $(-,+,+,+)$, and we use the real
(Majorana) representation of the gamma matrices $\ga_a$ (cf.~appendix
\ref{appgamma}). 
They satisfy $\{\ga_a,\ga_b\}=2\eta_{ab}$. We antisymmetrize with unit
weight, i.~e.~$\ga_{ab} \equiv \ga_{\left[a\right.}\ga_{\left.b\right]}
\equiv \frac{1}{2}[\ga_a,\ga_b]$ etc.
The parity matrix is $\ga_5 = \ga_{0123}$.}
\eqn
e^{-1}L &=& -\frac{1}{4}R + \frac{1}{2}\bar{\psi}_m\ga^{mnp}D_n\psi_p
            + \frac{1}{4}F_{mn}F^{mn} + \frac{i}{8}(F^{mn} + \hat{F}^{mn})
            \bar{\psi}_p\ga_{\left[m\right.}\ga^{pq}\ga_{\left.n\right]}
            \psi_q \nonumber \\
        &&  - \frac{1}{2\ell}\bar{\psi}_m\ga^{mn}\psi_n - \frac{3}{2\ell^2}.
            \label{lagrange}
\feqn
We see that the cosmological constant is $\Lambda = -3\ell^{-2}$.
$D_m$ denotes the
gauge- and Lorentz covariant derivative defined by
\eq
D_m = \nabla_m - i\ell^{-1}A_m, \label{gaugecovder}
\feq
where $\nabla_m$ is the Lorentz-covariant derivative
\eq
\nabla_m = \partial_m + \frac{1}{4}\om_m^{\;\;ab}\ga_{ab}.
\feq
The equation of motion for the spin connection $\om_m^{\;\;ab}$ reads
\eq
\om_{mab} = \Om_{mab} - \Om_{mba} - \Om_{abm},
\feq
where
\eq
\Om_{mn}^{\;\;\;\;a} \equiv \partial_{\left[m\right.}e_{\left.n\right]}^{\;a}
                        - \frac{1}{2}{\mathrm{Re}}(\bar{\psi}_m\ga^a\psi_n).
\feq
$\hat{F}_{mn}$ denotes the supercovariant field strenght given by
\eq
\hat{F}_{mn} = F_{mn} - {\mathrm{Im}}(\bar{\psi}_m\psi_n). \label{supcovfs}
\feq
The action is invariant under the local supertransformations
\eqn
\de e_m^{a} &=& {\mathrm{Re}}(\bar{\epsilon}\ga^a\psi_m), \nonumber \\
\de A_m &=& {\mathrm{Im}}(\bar{\epsilon}\psi_m), \label{transfsusy}\\
\de \psi_m &=& \hat{\nabla}_m \epsilon. \nonumber 
\feqn
In (\ref{transfsusy}) $\epsilon$ is an infinitesimal Dirac spinor, and
$\hat{\nabla}_m$ is the supercovariant derivative defined by
\eq
\hat{\nabla}_m = D_m + \frac{1}{2\ell}\ga_m + \frac{i}{4}\hat{F}_{ab}\ga^{ab}
                 \ga_m. \label{supcovder}
\feq
The supersymmetry algebra of gauged $N=2$ supergravity is $\mathrm{osp(4|2)}$.
It has the ten bosonic generators $M_{ab}, M_{a4}$ $(a=0,1,2,3)$ of the
AdS subalgebra $\mathrm{so(3,2)}$, two fermionic generators
$Q_{\alpha}^i$ $(i=1,2)$,
plus one additional bosonic generator
of $\mathrm{SO(2)}$ transformations, rotating the two supersymmetries
into each other.
The basic anticommutator is
\eq
\{Q^i_{\alpha},Q^j_{\beta}\} = \de^{ij}\left((\ga^aM_{a4} + i\ga^{ab}M_{ab})
                               C\right)_{\alpha\beta} + i(C_{\alpha\beta}Q_e
                               + i(C\ga^5)_{\al\be}Q_m)\epsilon^{ij}.
\feq
Here $C$ denotes the charge conjugation matrix, $Q_e$ and $Q_m$ are central
charges, and $\epsilon^{ij}$ is the permutation symbol in two dimensions.\\
Let us now return to the lagrangian (\ref{lagrange}).
As we are interested in the bosonic sector, we set $\psi_m=0$.
The field equations following from (\ref{lagrange}) are then the
Einstein-Maxwell equations with negative cosmological constant, thus
the black hole spacetimes discussed in the previous section represent
possible background solutions of gauged $N=2$ supergravity.
Invariance of these background solutions under the
supertransformations (\ref{transfsusy}) yields the equation for
Killing spinors
\eq
\hat{\nabla}_m \epsilon = 0. \label{killing}
\feq
The integrability conditions for (\ref{killing}) read
\eq
\hat{R}_{mn}\epsilon = 0, \label{integr}
\feq
where
\eq
\hat{R}_{mn} = \left[\hat{\nabla}_m,\hat{\nabla}_n\right] \label{supercurv}
\feq
is the supercurvature. (\ref{integr}) is necessary, but not sufficient
for the existence of Killing spinors. It assures that they exist locally,
but globally there may exist no Killing spinor due to topological reasons.
In the following sections,
we shall solve (\ref{killing}) and (\ref{integr}) for the
black hole spacetimes introduced in \ref{topbh}.

\section{Supersymmetry of Static Topological Black Holes} \label{nonrot}
\subsection{Integrability Conditions}
Let us first consider the static black hole spacetimes (\ref{NonRot}),
whose event horizons are Riemann surfaces of genus $g \ge 1$.
Setting $a=0$ in the spin connections given in the appendices
\ref{approttorus} and \ref{appHG}, one finds the supercovariant derivatives
\eqn
\hat{\nabla}_t &=& \partial_t - \frac{i}{\ell}\frac{q_e}{r} + \frac{1}{2\ell}
                   \sqrt{V(r)}\ga_0 + \frac{i}{4}F_{ab}\ga^{ab}\sqrt{V(r)}\ga_0
                   + \frac{1}{2r}\left(\frac{\eta}{r} - \frac{z^2}{r^2} +
                   \frac{r^2}{\ell^2}\right)\ga_{01}, \nonumber \\
\hat{\nabla}_r &=& \partial_r + \frac{1}{2\ell}\sqrt{V(r)}^{-1}\ga_1 +
                   \frac{i}{4}F_{ab}\ga^{ab}\sqrt{V(r)}^{-1}\ga_1, \nonumber \\
\hat{\nabla}_x &=& \partial_x - \frac{1}{2}\sqrt{V(r)}\ga_{12} +
                   \frac{1}{2\ell}r\ga_2 +
                   \frac{i}{4}rF_{ab}\ga^{ab}\ga_2, \nonumber \\
\hat{\nabla}_y &=& \partial_y - \frac{1}{2}\sqrt{V(r)}\ga_{13} 
                   - \frac{i}{\ell}q_m x + \frac{1}{2\ell}
                   r\ga_3 + \frac{i}{4}rF_{ab}\ga^{ab}\ga_3
\feqn
for toroidal topology (we consider here only the case $\tau = i$, $\tau$
denoting the Teichm\"uller parameter introduced in section
\ref{topbh}), and
\eqn
\hat{\nabla}_t &=& \partial_t - \frac{i}{\ell}\frac{q_e}{r} + \frac{1}{2\ell}
                   \sqrt{V(r)}\ga_0 + \frac{i}{4}F_{ab}\ga^{ab}\sqrt{V(r)}\ga_0
                   + \frac{1}{2r}\left(\frac{\eta}{r} - \frac{z^2}{r^2} +
                   \frac{r^2}{\ell^2}\right)\ga_{01}, \nonumber \\
\hat{\nabla}_r &=& \partial_r + \frac{1}{2\ell}\sqrt{V(r)}^{-1}\ga_1 +
                   \frac{i}{4}F_{ab}\ga^{ab}\sqrt{V(r)}^{-1}\ga_1, \nonumber \\
\hat{\nabla}_{\theta} &=& \partial_{\theta} - \frac{1}{2}\sqrt{V(r)}\ga_{12} +
                          \frac{1}{2\ell}r\ga_2 + \frac{i}{4}rF_{ab}
                          \ga^{ab}\ga_2, \\
\hat{\nabla}_{\phi} &=& \partial_{\phi} - \frac{1}{2}\sqrt{V(r)}\ga_{13}
                        \sinh\theta - \frac{1}{2}\ga_{23}\cosh\theta
                        - \frac{i}{\ell}q_m \cosh\theta + \frac{1}{2\ell}
                        r\ga_3\sinh\theta +
                        \frac{i}{4}rF_{ab}\ga^{ab}\ga_3\sinh\theta \nonumber
\feqn
for the higher genus ($g>1$) case. We recall that $V(r)$ is given by
\eq
V(r) = \de_{g,1} - 1 - \frac{2\eta}{r} + \frac{r^2}{\ell^2} + \frac{z^2}{r^2} .
\feq
One verifies that the supercurvature (\ref{supercurv}), like in the
Reissner-Nordstr\"om-AdS case studied by Romans \cite{romans},
can be written as a product
\eq
\hat{R}_{mn} = {\cal P}{\cal G}_{mn}(r){\cal O},
\feq
where ${\cal G}_{mn}(r)$ is $\ga_{mn}$ times some scalar-valued
function of $r$,
\eq
{\cal P} \equiv \frac{r^2}{2z}iF_{ab}\ga^{ab}\ga_1 \label{idem}
\feq
is an idempotent (and hence non-singular) operator, and $\cal O$ is
given by
\eq
{\cal O} = \sqrt{V(r)} + \frac{r}{\ell}\ga_1 + \left(\frac{z}{r} -
           \frac{\eta}{z}\right){\cal P}. \label{O}
\feq
As ${\cal G}_{mn}(r)$ and $\cal P$ are non-singular
for $z \neq 0$ (the case $z=0$
has to be considered separately), the integrability conditions for
Killing spinors are equivalent to the vanishing of $\det {\cal O}$.
We obtain for the determinant
\eq
\det {\cal O} = \left\{\de_{g,1} - 1 - \frac{2q_m}{\ell} - \frac{(\eta^2 -
                2q_m\eta r \ell^{-1})}{z^2}\right\}
                \left\{\de_{g,1} - 1 + \frac{2q_m}{\ell} - \frac{(\eta^2 +
                2q_m\eta r \ell^{-1})}{z^2}\right\}. \label{det}
\feq
$\det {\cal O}$ is a function of $r$, and for supersymmetric configurations,
this function must vanish identically. For genus $g=1$, this is fulfilled
in two cases. The first one appears for
\eq
q_m=0=\eta, \qquad \Rightarrow \quad V(r) = \frac{r^2}{\ell^2} +
\frac{q_e^2}{r^2}, \label{bogtor}
\feq
and the second one for
\eq
\eta=0 \; \wedge \; \ell = \infty \;\; (\Lambda = 0) \qquad \Rightarrow
\quad V(r) = \frac{z^2}{r^2}.
\feq
We observe that the lapse function is always positive in these cases,
i.~e.~the corresponding spacetimes represent naked singularities. This
result is very similar to the Reissner-Nordstr\"om-AdS black hole
considered in \cite{romans}, where the supersymmetric configurations are
also naked singularities.\\
For genus $g>1$, (\ref{det}) vanishes only for $\eta=0$ and $q_m = \pm \ell/2$.
This yields
\eq
V(r) = \left(\frac{r}{\ell} - \frac{\ell}{2r}\right)^2 + \frac{q_e^2}{r^2}.
\label{lapse}
\feq
For vanishing electric charge, spacetime (\ref{NonRot}), with $V(r)$
given by (\ref{lapse}), describes an extremal black hole. Thus, unlike the
case of spherical or toroidal event horizons, now we can hope to get a
supersymmetric static extremal black hole. Clearly, this is not obvious,
because it could be possible that the Killing spinors (which exist
locally, as (\ref{integr}) is satisfied), are not compatible with
the identifications which have to be carried out to get a compact
event horizon. We shall see
below, when we will construct explicitely the Killing spinors, that they
depend only on the radial coordinate $r$, and consequently they do respect the
identifications performed in the $(\theta,\phi)$-sector.\\
The extremal black hole found above is a solitonic object in the sense that
the limit $\ell^{-1} \to 0$ (we recall that $\ell^{-1}$ is the coupling
constant of the gauged theory, coupling the photon to the gravitini),
does not exist.\\
For genus $g>1$, there is still another case in which the integrability
conditions (\ref{integr}) are fulfilled, namely for $\eta = 0 = q_m = q_e$.
The spacetime is then a quotient space of AdS, and therefore locally
admits Killing spinors. However, we will see that they do not exist
globally, as the above mentioned identifications are not respected.
Thus the corresponding black hole is not supersymmetric.

\subsection{Killing Spinors}
We now turn to the issue of solving the Killing spinor equation
(\ref{killing}) explicitely for the diverse cases found above.

\subsubsection{Genus $g=1$, $V(r) = \frac{r^2}{\ell^2} + \frac{q_e^2}{r^2}$}
The spacetime describes an electrically charged naked singularity with
topology $\R^2 \times S^1 \times S^1$, and is asymptotically AdS.
Using the integrability condition ${\cal O}\epsilon = 0$, (\ref{killing})
simplifies in this case to
\eqn
\hat{\nabla}_r \epsilon &=& \left(\partial_r + \frac{1}{2\ell}\sqrt{V(r)}^{-1}
                            \ga_1 - \frac{iq_e}{2r^2}\sqrt{V(r)}^{-1}\ga_0
                            \right)\epsilon  = 0, \nonumber \\
\hat{\nabla}_t \epsilon &=& \partial_t \epsilon = 0, \nonumber \\
\hat{\nabla}_x \epsilon &=& \partial_x \epsilon = 0, \nonumber \\
\hat{\nabla}_y \epsilon &=& \partial_y \epsilon = 0.
\feqn
The Killing spinors thus depend only on $r$. One verifies that
\eq
{\cal Q} \equiv \frac{1}{2\sqrt{V(r)}}{\cal O} = \frac{1}{2}\left(1 +
                \frac{r}{\ell\sqrt{V(r)}}\ga_1 + \frac{q_e}{r\sqrt{V(r)}}
                \ga_0\right) 
\feq
is a projection operator. In the appendix of \cite{romans} one finds
the solution of the spinorial differential equation
\eq
\partial_r \epsilon(r) = (a(r) + b(r)\Ga_1 + c(r)\Ga_2)\epsilon(r),
\feq
where
\eq
(\Ga_1)^2 = (\Ga_2)^2 = 1, \qquad \{\Ga_1,\Ga_2\} = 0,
\feq
and $\epsilon(r)$ obeying the constraint
\eq
{\cal Q}\epsilon = 0,
\feq
with a projector ${\cal Q}$ given by
\eq
{\cal Q} = \frac{1}{2}(1 + \xi(r)\Ga_1 + \zeta(r)\Ga_2),
\feq
and
\eq
\xi^2 + \zeta^2 = 1, \qquad \zeta \neq 0.
\feq
In our case, this solution reads
\eq
\epsilon(r) = \left(\sqrt{\sqrt{V(r)} + \frac{r}{\ell}} + i\sqrt{\sqrt{V(r)}
              - \frac{r}{\ell}}\ga_0\right)P(-\ga_1)\epsilon_0,
\label{killspin1}
\feq
where $\epsilon_0$ is a constant spinor, and
\eq
P(-\ga_1) \equiv \frac{1}{2}(1-\ga_1)
\feq
is another projection operator, which reduces the complex dimension of
the space of Killing spinors from four to two. If the electric charge
also vanishes, the spacetime is simply a quotient of AdS,
representing the background (zero Hawking temperature) of uncharged
toroidal black holes. Then
the operators $\cal P$ (\ref{idem}) and $\cal O$ (\ref{O}) are
ill-defined, so
we must consider this case separately. It is clear that locally as many
Killing spinors as in AdS exist (four complex-dimensional solution space),
but we find that the only ones respecting the identifications one
carries out to compactify the $(x,y)$-sector to a torus,
are those resulting from (\ref{killspin1}) by
setting $q_e=0$, i.~e.
\eq
\epsilon(r) = \sqrt{r}P(-\ga_1)\epsilon_0,
\feq
so we have again a two complex-dimensional space of Killing spinors.

\subsubsection{Genus $g=1$, $V(r) = \frac{z^2}{r^2}$}
This space is not asymptotically AdS and represents a dyonic naked singularity
with topology $\R^2 \times S^1 \times S^1$.
For completeness we give the Killing spinors also for this case. Making use
of ${\cal O}\epsilon = 0$, (\ref{killing}) reduces to
\eqn
\hat{\nabla}_r \epsilon &=& \left(\partial_r + \frac{1}{2r}
                            \right)\epsilon  = 0, \nonumber \\
\hat{\nabla}_t \epsilon &=& \partial_t \epsilon = 0, \nonumber \\
\hat{\nabla}_x \epsilon &=& \partial_x \epsilon = 0, \nonumber \\
\hat{\nabla}_y \epsilon &=& \partial_y \epsilon = 0.
\feqn
Taking into account the constraint ${\cal O}\epsilon = 0$, one obtains
the solution
\eq
\epsilon(r) = r^{-1/2}P(-i\ga_0\frac{q_e}{z}+i\ga_{123}\frac{q_m}{z})
              \epsilon_0.
\feq
Again, a projection operator
\eq
P(-i\ga_0\frac{q_e}{z}+i\ga_{123}\frac{q_m}{z}) \equiv
\frac{1}{2}(1-i\ga_0\frac{q_e}{z}+i\ga_{123}\frac{q_m}{z})
\feq
acts on $\epsilon_0$, reducing the dimension of the solution space
from four to two.

\subsubsection{Genus $g>1$, $V(r) = \left(\frac{r}{\ell} - \frac{\ell}{2r}
\right)^2 + \frac{q_e^2}{r^2}$}
We now focus our attention on the case of spacetime topology
$\R^2 \times S_g$, $S_g$ being a Riemann surface of genus $g>1$.
The mass parameter $\eta$ is zero, and the magnetic charge $q_m$ equals
$\pm \ell/2$. We shall consider only $q_m = + \ell/2$, the other sign
giving identical results.
For nonvanishing electric charge, we have a dyonic naked
singularity; for $q_e=0$, however, we get an extremal magnetically
charged black hole
(a magnetic monopole hidden by an event horizon having the topology
of a Riemann surface).\\
In this case the operator $\cal O$ is not proportional to a projector,
but rather is a linear combination of two projection operators
$P(-i\ga_{23}) \equiv (1-i\ga_{23})/2$ \footnote{from now on, with
$P({\cal L})$, where $\cal L$ is an operator, we always intend
$(1 + {\cal L})/2$.}
and
\eq
{\cal Q} \equiv \frac{1}{2} + \frac{1}{2\sqrt{V(r)}}\left\{\left(\frac{r}{\ell}
                - \frac{\ell}{2r}\right)\ga_1 + i\ga_0\frac{q_e}{r}\right\}.
\feq
We find
\eq
{\cal O} = 2\sqrt{V(r)}{\cal Q} + \frac{\ell}{r}\ga_1P(-i\ga_{23})
\feq
and
\eqn
-\frac{1}{2}\left[\sqrt{V(r)} - \frac{r}{\ell}\ga_1 - \frac{1}{r}(i\ga_0q_e -
\frac{\ell}{2}i\ga_{123})\right]{\cal O} &=& P(-i\ga_{23}), \\
\frac{\ell}{4\sqrt{V(r)}r}\ga_1\left[\sqrt{V(r)} + \frac{r}{\ell}\ga_1 -
\frac{1}{r}(i\ga_0q_e - \frac{\ell}{2}i\ga_{123})\right]{\cal O} &=&
{\cal Q}, \\
{[}{\cal Q}, P(-i\ga_{23})] &=& 0.
\feqn
The integrability condition ${\cal O}\epsilon = 0$ is thus equivalent to
the two conditions
\eq
P(-i\ga_{23})\epsilon = 0, \qquad {\cal Q}\epsilon = 0.
\feq
The Killing spinor equations then simplify to
\eqn
\hat{\nabla}_r \epsilon &=& \left(\partial_r + \frac{1}{2r} + \frac{1}{\ell
                            \sqrt{V(r)}}\ga_1
                            \right)\epsilon  = 0, \nonumber \\
\hat{\nabla}_t \epsilon &=& \partial_t \epsilon = 0, \nonumber \\
\hat{\nabla}_{\theta} \epsilon &=& \partial_{\theta} \epsilon = 0, \nonumber \\
\hat{\nabla}_{\phi} \epsilon &=& \partial_{\phi} \epsilon = 0.
\feqn
The solution of the radial equation can again be constructed using the
appendix of \cite{romans}, yielding
\eq
\epsilon(r) = \left(\sqrt{\sqrt{V(r)} + \frac{r}{\ell} - \frac{\ell}{2r}} -
              \sqrt{\sqrt{V(r)} - \frac{r}{\ell} + \frac{\ell}{2r}}i\ga_0
              \right)P(-\ga_1)P(i\ga_{23})\epsilon_0.
\feq
Now the constant spinor $\epsilon_0$ is subject to a double projection,
so the solution space is one (complex)-dimensional. The Killing spinor
does not depend on the coordinates $\theta$, $\phi$, thus it respects
the identifications we have done in the $(\theta, \phi)$-sector to
obtain a Riemann surface. Hence, for zero electric charge,
we have obtained a supersymmetric
extremal static black hole. This was not possible for spherical
event horizons, i.~e.~for the Reissner-Nordstr\"om-AdS black hole,
where all supersymmetric configurations were naked singularities
\cite{romans}. So we see that admitting other spacetime topologies
changes the supersymmetry properties.\\
According to (\ref{mass}) and (\ref{magncharge}), the mass and the
magnetic charge of the supersymmetric higher genus black holes
considered above, are given by $M=\ell(g-1)/3\sqrt{3}$ and $Q_m=\ell(g-1)/2$,
i.~e.~we have
\eq
M^2 = \frac{4}{27}Q_m^2
\feq
as Bogomol'nyi bound. This bound supports the view advocated in
\cite{vanzo}, namely that the mass of the higher genus black holes
is not simply given by the parameter $\eta$ appearing in (\ref{lapsegen}),
but rather by (\ref{mass}), i.~e.~the background which has to be
subtracted in the mass calculation, is not simply the one with $\eta=0$.
Note that in \cite{vanzo}, this conclusion emerges from thermodynamical
considerations, and has nothing to do with supersymmetry. We found here
that also supersymmetry as an independent argument supports this
point of view.

\subsubsection{Genus $g>1$, $V(r) = -1 + \frac{r^2}{\ell^2}$}
This is a quotient space of AdS describing an uncharged black hole.
Without identifications in the $(\theta, \phi)$-sector the spacetime
is simply AdS viewed by a uniformely accelerated observer, the
(noncompact) horizon being its acceleration horizon \cite{vanzo}.
Only the compactification of the surfaces of constant $r$ and $t$
makes the spacetime to become a black hole, with the singularity at
$r=0$ being a causal one, i.~e.~the manifold cannot be continued
beyond this singularity, otherwise one would have closed timelike curves
\cite{amin}. It is clear that locally this spacetime admits as many
Killing spinors as AdS, but we have to check if they respect the
identifications. The Killing spinor equations read
\eqn
\hat{\nabla}_r \epsilon &=& \left(\partial_r + \frac{1}{2\ell\sqrt{V(r)}}\ga_1
                            \right)\epsilon  = 0, \nonumber \\
\hat{\nabla}_t \epsilon &=& \left(\partial_t + \frac{r}{2\ell^2}\ga_{01} +
                            \frac{1}{2\ell}\sqrt{V(r)}\ga_0\right)
                            \epsilon = 0, \nonumber \\
\hat{\nabla}_{\theta} \epsilon &=& \left(\partial_{\theta} + \frac{r}{2\ell}
                                   \ga_2 - \frac{1}{2}\sqrt{V(r)}\ga_{12}
                                   \right)\epsilon = 0, \nonumber \\
\hat{\nabla}_{\phi} \epsilon &=& \left(\partial_{\phi} + \frac{1}{2}\ga_3
                                 (\frac{r}{\ell} + \sqrt{V(r)}
                                 \ga_1)\sinh \theta -
                                 \frac{1}{2}\ga_{23}\cosh\theta
                                 \right)\epsilon = 0.
\feqn
The solution is
\eqn
\lefteqn{\epsilon(r,t,\theta,\phi) =} \nonumber \\
          & & (\sqrt{\frac{r}{\ell}+1} - \sqrt{\frac{r}{\ell}-1}\ga_1)
              (\cosh\frac{t}{2\ell} - \ga_{01}\sinh\frac{t}{2\ell})
              (\cosh\frac{\theta}{2} - \ga_2\sinh\frac{\theta}{2})
              (\cos\frac{\phi}{2} + \ga_{23}\sin\frac{\phi}{2})\epsilon_0.
\feqn
As an explicit $\phi$-dependence appears, the Killing spinors are not
invariant under the transformations of the discrete group used
in the identifications,
and the black hole is not supersymmetric. Clearly
this was to be expected, as a supersymmetric black hole necessarily
must have zero temperature (note, however, that the converse is not
true in general), whereas the hole considered above has nonvanishing
Hawking temperature $T=1/2\pi\ell$.\\
Note that the minimal coupling of the photon and the gravitini in the
action of gauged $N=2$ supergravity gives rise to a Dirac quantization
of the magnetic charge. In the spherical static case, this condition is
automatically fulfilled for the supersymmetric solutions \cite{romans}.
Finding the Dirac quantization condition in presence of unusual topologies,
as is partially the case here, is a nontrivial task,
which involves $\mathrm{U}(1)$-bundles over
Riemann surfaces of genus $g \ge 1$. We will discuss this
problem in a forthcoming publication.

\section{Supersymmetry of Rotating AdS Black Holes} \label{rot}
Now we turn to the rotating generalizations of the static black holes
considered above. As the metrics are rather complicated, the investigation of
supersymmetry properties of these spacetimes is a quite formidable task.
However, as we shall see below, it is still possible to solve
explicitely the integrability conditions, yielding some interesting results.

\subsection{Cylindrical Event Horizons}
Let us first consider the black hole spacetime (\ref{rottorus}).
The supercovariant derivatives read
\eqn
\hat{\nabla}_t &=& \partial_t + \frac{1}{2}\om_t^{\;\;01}\ga_{01} +
                   \frac{1}{2}\om_t^{\;\;23}\ga_{23} + i\frac{q_er+q_mPa}{\ell
                   \rho^2} + \left(\frac{1}{2\ell\rho} + \frac{i}{4\rho}
                   F_{ab}\ga^{ab}\right)(\sqrt{\Delta_r}\ga_0 +
                   \sqrt{\Delta_P}a\ga_3), \nonumber \\
\hat{\nabla}_r &=& \partial_r + \frac{1}{2}\om_r^{\;\;12}\ga_{12} +
                   \frac{1}{2}\om_r^{\;\;03}\ga_{03} + \frac{\rho}{2\ell
                   \sqrt{\Delta_r}}\ga_1 + \frac{i\rho}{4\sqrt{\Delta_r}}
                   F_{ab}\ga^{ab}\ga_1, \nonumber \\
\hat{\nabla}_P &=& \partial_P + \frac{1}{2}\om_P^{\;\;12}\ga_{12} +
                   \frac{1}{2}\om_P^{\;\;03}\ga_{03} + \frac{\rho}{2\ell
                   \sqrt{\Delta_P}}\ga_2 + \frac{i\rho}{4\sqrt{\Delta_P}}
                   F_{ab}\ga^{ab}\ga_2, \nonumber \\
\hat{\nabla}_{\phi} &=& \partial_{\phi} + \frac{1}{2}\om_{\phi}^{\;\;01}
                        \ga_{01} + \frac{1}{2}\om_{\phi}^{\;\;23}\ga_{23} + 
                        \frac{1}{2}\om_{\phi}^{\;\;02}\ga_{02} +
                        \frac{1}{2}\om_{\phi}^{\;\;13}\ga_{13} +
                        i\frac{q_eraP^2-q_mPr^2}{\ell\rho^2} \nonumber \\
                     && + \left(\frac{1}{2\ell\rho} + \frac{i}{4\rho}
                        F_{ab}\ga^{ab}\right)(\sqrt{\Delta_r}aP^2\ga_0 -
                        \sqrt{\Delta_P}r^2\ga_3),
\feqn
with the electromagnetic field $F_{ab}$ (\ref{fieldrottorus}), and the
spin connection $\om_m^{\;\;ab}$ given in appendix
\ref{approttorus}. Similarly to the nonrotating case, we find for the
supercurvature
\eq
\hat{R}_{mn} = {\cal P}{\cal G}_{mn}(r,P){\cal O},
\feq
with the idempotent operator $\cal P$ now given by
\eq
{\cal P} \equiv \frac{\rho^2}{2z}iF_{ab}\ga^{ab}\ga_1, \label{idemrot}
\feq
and
\eqn
{\cal O} &=& \frac{1}{\rho^2}\left\{-\sqrt{\Delta_P}ra\ga_{03} -
             \sqrt{\Delta_r}Pa\ga_{0123} + \sqrt{\Delta_r}r +
             \sqrt{\Delta_P}a^2P\ga_{12}\right\} + \frac{\rho}{\ell}\ga_1
             \nonumber \\
          && + \frac{1}{\rho^3}\left\{-aP\left[\frac{\eta}{z}(3r^2-a^2P^2)
             -2rz\right]\ga_{0123} + \left[\frac{\eta r}{z}(r^2-3a^2P^2) +
             z(a^2P^2-r^2)\right]\right\}{\cal P}.
\feqn
${\cal P}$ and ${\cal G}_{mn}(r,P)$ are in general nonsingular,
so the integrability condition is again $\det {\cal O} = 0$. The determinant
reads
\eqn
\det {\cal O} &=& \frac{1}{\ell^2z^4}\left[(\ell^2\eta^4 - 4z^4(q_m^2+a^2)) +
                  8q_m^2z^2\eta r - 4q_m^2\eta^2r^2\right. \nonumber \\
               && \left. + 8a\eta q_m q_e z^2 P +
                  4a^2q_m^2\eta^2P^2 - 8a\eta^2q_mq_ePr\right].
\feqn
For $z \neq 0$ (which we presupposed, as otherwise $\cal P$ is ill-defined),
the requirement that $\det {\cal O}$ be vanishing identically as a function
of $r$ and $P$, yields the conditions
\eqn
q_m\eta &=& 0, \nonumber \\
\ell^2\eta^4 &=& 4z^4(q_m^2+a^2).
\feqn
The case $\eta=0$ is not of particular interest for us, as it does not
represent a black hole spacetime. Therefore we will assume $\eta > 0$,
from which follows $q_m=0$ and
\eq
\eta^2 = \frac{2a}{\ell}q_e^2 \label{bogcyl}
\feq
for supersymmetric configurations. We observe that for $a=0$,
(\ref{bogcyl}) reduces correctly to (\ref{bogtor}), i.~e.~$\eta=0$.
We do not construct the Killing spinors
explicitely for the rotating case. However, we have seen that for $a=0$
there is a two-dimensional solution space of Killing spinors depending
only on the radial coordinate $r$ (cf.~(\ref{killspin1})),
and we expect a similar behaviour also
for the rotating black holes, especially we expect the Killing spinors
not to depend on the angular coordinate $\phi$, and hence to respect
the identification $\phi \sim \phi + 2\pi$, leading to cylindrical topology.\\
(\ref{bogcyl}) can be compared with the extremality condition
\eq
a^2+q_e^2 = \frac{3\eta^{4/3}\ell^{2/3}}{2^{4/3}}.
\feq
Combining this with (\ref{bogcyl}), we obtain the relation
\eq
a^2 + q_e^2 = \frac{3}{2^{2/3}}q_e^{4/3}a^{2/3}. \label{hom}
\feq
This is a homogenous equation, the solutions are therefore on straight
lines $q_e^2 = \beta^2 a^2$, with $\beta > 0$. Inserting this into
(\ref{hom}), one determines $\beta = 2$. Using the supersymmetry
condition (\ref{bogcyl}) finally yields that extremal supersymmetric
rotating cylindrical black holes are parametrized by
\eqn
|q_e| &=& \frac{\ell^{1/3}\eta^{2/3}}{2^{1/6}}, \nonumber \\
a = \frac{|q_e|}{\sqrt{2}} &=& \frac{\ell^{1/3}\eta^{2/3}}{2^{2/3}}.
\feqn
We have obtained an interesting result: In order to get extremal
supersymmetric black holes with cylindrical event horizon topology, we must
allow the holes to carry
angular momentum, for in the static case all the supersymmetric configurations
are naked singularities. This behaviour is similar to that of the
spherical Reissner-Nordstr\"om-AdS solution, for which Kosteleck\'{y} and
Perry showed by considering the Bogomol'nyi bound arising from the
supersymmetry algebra, that solitonic black holes must be rotating
\cite{perry}.

\subsection{Generalization of the Higher Genus Case}

We turn now to the rotating generalizations (\ref{HG}) of the
higher genus black hole spacetimes, i.~e.~to the rotating charged
black membranes in AdS space.
For the supercovariant
derivatives, one gets
\eqn
\hat{\nabla}_t &=& \partial_t + \frac{1}{2}\om_t{}^{01}\ga_{01} +
                   \frac{1}{2}\om_t{}^{23}\ga_{23} + i\frac{q_er+q_ma\cosh
                   \theta}{\ell\Xi\rho^2}\nonumber\\ 
                   && + \left(\frac{1}{2\ell\Xi\rho} 
                   +\frac{i}{4\Xi\rho}F_{ab}\ga^{ab}\right)
                   (\sqrt{\Delta_r}\ga_0+\sqrt{\Delta_\theta}a\sinh\theta\ga_3)
                   , \nonumber \\
\hat{\nabla}_r &=& \partial_r + \frac{1}{2}\om_r{}^{03}\ga_{03} +
                   \frac{1}{2}\om_r{}^{12}\ga_{12} + \frac{\rho}{2\ell
                   \sqrt{\Delta_r}}\ga_1 + \frac{i\rho}{4\sqrt{\Delta_r}}
                   F_{ab}\ga^{ab}\ga_1, \nonumber \\
\hat{\nabla}_\theta &=& \partial_\theta + \frac{1}{2}\om_\theta{}^{03}\ga_{03}+
                   \frac{1}{2}\om_\theta{}^{12}\ga_{12} + \frac{\rho}{2\ell
                   \sqrt{\Delta_\theta}}\ga_2 + 
                   \frac{i\rho}{4\sqrt{\Delta_\theta}}F_{ab}\ga^{ab}\ga_2, 
                   \nonumber \\
\hat{\nabla}_{\phi} &=& \partial_{\phi} + \frac{1}{2}\om_{\phi}^{\;\;01}
                        \ga_{01} + \frac{1}{2}\om_{\phi}^{\;\;02}\ga_{02} + 
                        \frac{1}{2}\om_{\phi}^{\;\;13}\ga_{13} +
                        \frac{1}{2}\om_{\phi}^{\;\;23}\ga_{23} +
                        i\frac{q_era\sinh^2\theta-q_m(r^2+a^2)\cosh\theta}
                        {\ell\Xi\rho^2} \nonumber \\
                     && + {\sinh\theta\over2\Xi\rho}\left(\frac{1}{\ell}
                        +\frac{i}{2}
                        F_{ab}\ga^{ab}\right)(\sqrt{\Delta_r}a\sinh\theta\ga_0
                        -\sqrt{\Delta_\theta}(r^2+a^2)\ga_3),
\feqn
with the electromagnetic field $F_{ab}$ given by Eq. (\ref{fieldHG}), and the
spin connection $\om_m^{\;\;ab}$ given in appendix
\ref{appHG}. Analogously to the previous cases, we find for the
supercurvature
\eq
\hat{R}_{mn} = {\cal P}{\cal G}_{mn}(r,\theta){\cal O},
\feq
with the idempotent operator $\cal P$ again given by Eq. \rif{idemrot},
and
\eqn
{\cal O} &=& {1\over\rho^2}\left\{-\sqrt{\Delta_\theta}ra\sinh\theta\ga_{03} -
             \sqrt{\Delta_r}a\cosh\theta\ga_{0123} + \sqrt{\Delta_r}r +
             \sqrt{\Delta_\theta}a^2\sinh\theta\cosh\theta\ga_{12}\right\} 
             +\frac{\rho}{\ell}\ga_1
             \nonumber \\
          && + \frac{1}{\rho^3}\left\{-a\cosh\theta
             \left[\frac{\eta}{z}(3r^2-a^2\cosh^2\theta)-2rz\right]\ga_{0123} 
             \right.\nonumber\\
          && +\left.\left[\frac{\eta r}{z}(r^2-3a^2\cosh^2\theta) +
             z(a^2\cosh^2\theta-r^2)\right]\right\}{\cal P}.
\feqn
Still ${\cal P}$ and ${\cal G}_{mn}(r,P)$ are in general nonsingular,
so the integrability condition is again $\det {\cal O} = 0$. The determinant
reads
\eqn
\det {\cal O} &=& \frac{1}{\ell^4z^4}
                  \left[((a^2+\ell^2)^2-4q_m^2\ell^2)
                  z^4+2\ell^2(\ell^2-a^2)\eta^2z^2
                  +\ell^4\eta^4+4\ell^2\eta q_m^2r(2z^2-\eta r)
                  \right.\nonumber\\
               && \left.+8a\ell^2\eta q_eq_m(z^2-\eta r)\cosh\theta
                  +4a^2\ell^2\eta^2q_m^2\cosh^2\theta\right].
\feqn
We assume $z \neq 0$ in order that $\cal P$ being well-defined. Then
the requirement that $\det {\cal O}$ be vanishing identically as a function
of $r$ and $\theta$, yields the conditions
\eqn
&&q_m\eta = 0, \nonumber \\
&&\eta^4+2\lp1-{a^2\over \ell^2}\rp\eta^2z^2+\lp\lp1+{a^2\over
  \ell^2}\rp^2-{4q_m^2\over \ell^2}\rp z^4 = 0.
\feqn
The case $q_m=0$ admits no solution, hence supersymmetry requires $\eta=0$,
from which it follows that
\eq
q_m^2={\ell^2\over 4}\lp1+{a^2\over \ell^2}\rp^2 \label{bogHG}
\feq
for supersymmetric configurations. This yields
\eq
\Delta_r = \left[{r^2\over \ell}-{\ell\over 2}\lp1-{a^2\over \ell^2}
           \rp\right]^2+q_e^2,
\label{lapserotsol}
\feq
which is a strictly positive function for $q_e^2>0$ and has a positive
double root for vanishing electric charge and $a<\ell$.
As long as there is an electric 
charge, these supersymmetric solutions represent a naked singularity. 
For $q_e=0$ and $a < \ell$, however, we obtain
a supersymmetric, magnetically charged, rotating, extremal
black membrane, that has a
solitonic interpretation.
We observe that for $a=0$, (\ref{bogHG}) reduces correctly to
the result (\ref{lapse}), and we have generalized this solution to a
one-parameter family of extremal supersymmetric solutions.\\
In conclusion, we have shown that there exists also a rotating
generalization of the
extremal supersymmetric magnetic black hole found in section \ref{nonrot}.
Besides, we saw that supersymmetry requires $\eta=0$, and that in order to
get supersymmetric black objects, also the electric charge must vanish.
It is interesting to compare this with the cylindrical topology considered
in the previous subsection, where the {\it magnetic charge} was required to
be zero.

\subsection{Revisitation of the Kerr-Newman-AdS Black Hole}

Now, we turn back to the Kerr-Newmann-AdS black hole (\ref{KNAdS}), that has
already been treated by Kosteleck\'y and Perry \cite{perry}, analyzing the
Bogomol'nyi bound arising from the superalgebra.
We shall reconsider the problem by solving the
integrability condition, and show that the supersymmetry conditions are
more restrictive than those found in \cite{perry}.
The supercovariant derivatives read
\eqn
\hat{\nabla}_t &=& \partial_t + \frac{1}{2}\om_t{}^{01}\ga_{01} +
                   \frac{1}{2}\om_t{}^{23}\ga_{23} + i\frac{q_er+q_ma\cos
                   \theta}{\ell\Xi\rho^2}\nonumber\\ 
                   && + \left(\frac{1}{2\ell\Xi\rho} 
                   +\frac{i}{4\Xi\rho}F_{ab}\ga^{ab}\right)
                   (\sqrt{\Delta_r}\ga_0+\sqrt{\Delta_\theta}a\sin\theta\ga_3)
                   , \nonumber \\
\hat{\nabla}_r &=& \partial_r + \frac{1}{2}\om_r{}^{03}\ga_{03} +
                   \frac{1}{2}\om_r{}^{12}\ga_{12} + \frac{\rho}{2\ell
                   \sqrt{\Delta_r}}\ga_1 + \frac{i\rho}{4\sqrt{\Delta_r}}
                   F_{ab}\ga^{ab}\ga_1, \nonumber \\
\hat{\nabla}_\theta &=& \partial_\theta + \frac{1}{2}\om_\theta{}^{03}\ga_{03}+
                   \frac{1}{2}\om_\theta{}^{12}\ga_{12} + \frac{\rho}{2\ell
                   \sqrt{\Delta_\theta}}\ga_2 + 
                   \frac{i\rho}{4\sqrt{\Delta_\theta}}F_{ab}\ga^{ab}\ga_2, 
                   \nonumber \\
\hat{\nabla}_{\phi} &=& \partial_{\phi} + \frac{1}{2}\om_{\phi}^{\;\;01}
                        \ga_{01} + \frac{1}{2}\om_{\phi}^{\;\;02}\ga_{02} + 
                        \frac{1}{2}\om_{\phi}^{\;\;13}\ga_{13} +
                        \frac{1}{2}\om_{\phi}^{\;\;23}\ga_{23} -
                        i\frac{q_era\sin^2\theta+q_m(r^2+a^2)\cos\theta}
                        {\ell\Xi\rho^2} \nonumber \\
                     && - {\sin\theta\over2\Xi\rho}\left(\frac{1}{\ell}
                        +\frac{i}{2}
                        F_{ab}\ga^{ab}\right)(\sqrt{\Delta_r}a\sin\theta\ga_0
                        +\sqrt{\Delta_\theta}(r^2+a^2)\ga_3),
\feqn
with the electromagnetic field $F_{ab}$ given by Eq. (\ref{fieldKNAdS}),
and the spin connection $\om_m^{\;\;ab}$ given in appendix
\ref{appKNAdS}. Again, we find for the
supercurvature
\eq
\hat{R}_{mn} = {\cal P}{\cal G}_{mn}(r,\theta){\cal O},
\feq
where, as usual, the idempotent operator $\cal P$ is defined by
\rif{idemrot}, and
\eqn
{\cal O} &=& {1\over\rho^2}\left\{\sqrt{\Delta_\theta}ra\sin\theta\ga_{03} -
             \sqrt{\Delta_r}a\cos\theta\ga_{0123} + \sqrt{\Delta_r}r -
             \sqrt{\Delta_\theta}a^2\sin\theta\cos\theta\ga_{12}\right\} 
             -\frac{\rho}{\ell}\ga_1
             \nonumber \\
          && + \frac{1}{\rho^3}\left\{-a\cos\theta
             \left[\frac{\eta}{z}(3r^2-a^2\cos^2\theta)-2rz\right]\ga_{0123} 
             \right.\nonumber\\
          && +\left.\left[\frac{\eta r}{z}(r^2-3a^2\cos^2\theta) +
             z(a^2\cos^2\theta-r^2)\right]\right\}{\cal P}.
\feqn
${\cal P}$ and ${\cal G}_{mn}(r,\theta)$ are in general nonsingular,
so the integrability condition is again $\det {\cal O} = 0$. The determinant
reads
\eqn
\det {\cal O} &=& \frac{1}{\ell^4z^4}
                  \left[((\ell^2-a^2)^2-4q_m^2\ell^2)z^4-2\ell^2
                  (\ell^2+a^2)m^2z^2
                  +\ell^4m^4+4\ell^2m q_m^2r(2z^2-m r)\right.\nonumber\\
               && \left.+8a\ell^2m q_eq_m(z^2-m r)\cos\theta
                  +4a^2\ell^2m^2q_m^2\cos^2\theta\right].
\feqn
For $z \neq 0$ (which we still assume, as otherwise $\cal P$ is ill-defined),
the requirement that $\det {\cal O}$ be vanishing identically as a function
of $r$ and $\theta$, yields the conditions
\eqn
&& m q_m = 0, \nonumber \\
&&m^4-2\lp1+{a^2\over \ell^2}\rp m^2z^2+\lp\lp1-{a^2\over
  \ell^2}\rp^2-{4q_m^2\over \ell^2}\rp z^4 = 0.\label{susyKNAdS}
\feqn
To solve the integrability conditions, we have to put either $q_m$ or $m$
to zero.
In the first case, $q_m=0$, and the second condition of \rif{susyKNAdS}
yields
\eq
m^2=\lp1\pm{a\over \ell}\rp^2q_e^2, \label{bogKNAdS}
\feq
and we have electrically charged possibly supersymmetric
configurations. In the limit case $a=0$ we recover the usual
condition $m^2=q_e^2$.\\
Now one may wonder why the constraint (\ref{bogKNAdS}) on the electric
charge cannot be rotated into a similar constraint on the magnetic charge
or a combination of the two by an electromagnetic duality transformation.
This is a legitimate question, since dualities of this kind normally
are also valid in supergravity theories \cite{ferrara} (see also \cite{tanii}
for a recent review), where a typical duality transformation is of the form
\eq
\delta \hat{F}^{mn} = \frac{1}{2}ie^{-1}\lambda (^{\ast}\hat{F})^{mn}.
\label{dual}
\feq
Here $\lambda$ is a real parameter and
$\hat{F}^{mn}$ denotes the supercovariant field strength (\ref{supcovfs}),
involving also fermion fields in addition to $F^{mn}$. (In general, also
the fermions have to be transformed, cf.~\cite{ferrara}).
In fact, the Bogomol'nyi bound arising in ungauged supergravities usually
involves electric and magnetic charges in a duality invariant way.
Now it is clear that invariances of the kind (\ref{dual})
can hold only if the vector fields
interact only through the field strength
$F^{mn}$ with the spinors of the theory,
like it is the case e.~g.~in the trilinear coupling of the fourth
term on the right-hand side of (\ref{lagrange}).
However, if one introduces a minimal coupling of the vector fields
to the fermions by the gauge potential $A_m$ like in (\ref{gaugecovder}),
electromagnetic duality invariance is broken, which means that
gauged supergravity theories cannot have the usual duality
symmetries present in the ungauged theories. Therefore one should not
be surprised if the supersymmetry conditions found above break this
invariance, and treat electric and magnetic charges in a different way.
In our case the bosonic sector of (\ref{lagrange}) is duality invariant,
hence performing a duality on the supersymmetric solution (\ref{bogKNAdS})
we obtain again a solution of the supergravity equations; however the
duality breaks the supersymmetry of the solution
because the Killing spinor equation is not
duality invariant, and the condition (\ref{killing}) does not hold anymore.\\
Inserting
\rif{bogKNAdS} into the extremality condition \rif{extrKNAdS}, we obtain
the relation
\eq
m^2=\ell a\lp 1+{a\over \ell}\rp^4;
\feq
the configurations that satisfy the latter equation are hence possibly
supersymmetric extremal black holes, which carry electric charge and rotate.
Expression \rif{bogKNAdS} essentially coincides with the Bogomol'nyi bound
found by Kosteleck\'y and Perry \cite{perry}, emerging from the susy algebra.
However, we stress
the fact that supersymmetry requires a vanishing magnetic charge in this
case; the authors of \cite{perry} missed this condition in their paper.
The reason for this is that the supersymmetry condition given in
\cite{perry} is necessary, but not sufficient. If one derives the
Bogomol'nyi bound from the superalgebra, one additionally has to satisfy
the Witten equation \cite{witten} (see also \cite{ghw,ghhp})
on a three-dimensional spacelike hypersurface $\Sigma$,
\eq
^{(4)}\hat{\nabla}_m\epsilon = 0, \label{witten}
\feq
in order to assure the existence of Killing spinors. 
Here $^{(4)}\hat{\nabla}_m$
is the projection into $\Sigma$ of the four-dimensional
supercovariant derivative $\hat{\nabla}_m$ (\ref{supcovder}), and 
$\epsilon$
is a spinor field obeying the fall-off condition
\eq
\hat{\nabla}_m \epsilon = O(\frac{1}{r^2}).
\feq
Now a priori
it is not evident that (\ref{witten}) possesses a solution
in the case under consideration (although a unique solution may
exist in simpler cases, cf.~\cite{witten,ghhp} for a discussion),
and we expect
that the condition for (\ref{witten}) to have a solution will be just the
vanishing of the magnetic charge.\\
Let us now return to the conditions (\ref{susyKNAdS}).
In the other case, $m=0$, we obtain a
supersymmetric solution with
\eq
q_m^2 = \frac{\ell^2}{4}\lp1-{a^2\over \ell^2}\rp^2, 
\feq
which describes supersymmetric naked singularities, as can be seen from the
function
\eq
\Delta_r = \left[{r^2\over \ell}+{\ell\over 2}\lp1+{a^2\over \ell^2}
           \rp\right]^2+q_e^2.
\feq
This solution was not obtained in \cite{perry}; it is the spherical
analogue of the rotating solitonic membrane desribed by \rif{lapserotsol}.
However, in the latter case we have an event horizon for $q_e=0$, whereas
for spherical topology the singularity is naked.\\
Summarized, we can state that in order to get extremal
supersymmetric Kerr-Newman-AdS black holes, we must
allow the holes to carry
angular momentum, for in the static case all the supersymmetric configurations
are naked singularities. Besides, the extremal supersymmetric holes
carry electric charge only, making the result of Kosteleck\'{y} and
Perry more precise.

\section{Summary and Discussion}\label{disc}

In the present paper we considered four-dimensional asymptotically
AdS dyonic black holes with various topologies in the context of gauged
$N=2$ supergravity. For toroidal or cylindrical topology, all
static configurations
preserving some amount of supersymmetry, are naked singularities,
a behaviour common from the spherical Reissner-Nordstr\"om-AdS
case studied previously by Romans. However, for black holes whose
event horizons are Riemann surfaces of genus $g>1$, we found an extremal
supersymmetric black hole carrying purely magnetic charge, and
admitting a one-(complex) dimensional solution space of Killing spinors.
As we have seen, this solitonic object possesses also a rotating
generalization, whose analogue
in the Kerr-Newman-AdS case represents a naked singularity.
However, for cylindrical or spherical topology,
extremal supersymmetric black holes carrying only electric charge,
can appear for nonvanishing angular momentum. Hence, in these cases, solitonic
black holes must rotate.
This is in agreement with
the Kerr-Newman-AdS result of \cite{perry}, emerging from considerations
of the superalgebra. Yet, the authors of \cite{perry} did not obtain
the condition of vanishing magnetic charge for these BPS states,
so we have made more precise the Bogomol'nyi bound found in \cite{perry}.
The rotating supersymmetric states with purely electric charge, appearing
for cylindrical or spherical topology, have no analogue
for the rotating generalizations of the higher genus solutions.\\
Summarized, we can state that admitting unusual black hole topologies,
and allowing the holes also to carry angular momentum, can lead to a new
variety of states preserving some supersymmetry. It would be very
interesting to understand the Bogomol'nyi bounds, found for
unusual topology, in terms of the superalgebra. However, this requires
a careful definition of the mass and angular momentum of these
rotating black
configurations. As such a definition is a rather delicate question \cite{KMV}
\footnote{In particular, for the rotating black membrane solution
(\ref{HG}), the total mass and angular momentum are infinite. If one
tries to define
conserved quantities per unit brane-volume, as it is
usually done for $p$-branes, they will depend on the coordinate $\theta$
on the membrane, and hence they are not constant.},
the mentioned enterprise becomes a nontrivial
task, which we leave for future investigations.\\
All the metrics considered in this paper, are special cases of the
most general known Petrov type D metric found by Plebanski and Demianski
\cite{plebdem}, and probably this metric can lead to other black configurations
hitherto unknown. Therefore
it would also be interesting to investigate, under which conditions
this most general type D metric admits Killing spinors.\\
Another future line of research would be to look for similar
supersymmetric black hole
solutions with unusual topology in the context of gauged $N=4$ supergravity
\cite{freedschwarz}. As was shown recently by Chamseddine \cite{cham} and
Chamseddine and Volkov \cite{chamvolk}, gauged $N=4$
$\mathrm{SU(2)}\times\mathrm{SU(2)}$ supergravity
in four dimensions can be
obtained by compactifying $N=1$ supergravity in ten dimensions on the
group manifold $S^3 \times S^3$, and hence can also be obtained by
compactifying $N=1$ supergravity in eleven dimensions, which is
the low energy limit of M-theory, the most promising candidate for
a theory unifying gravity with the other fundamental interactions.
This connection would it make possible to lift the supersymmetric
black hole solutions of gauged $N=4$ supergravity to ten or eleven
dimensions, and thus to regard them as BPS solutions of string theory
or M-theory. Viewed in this larger context, the black holes eventually
would be accessible to a microscopic interpretation of their entropy,
using the tools of string-/M-theory, like D-brane technology.

\section*{Acknowledgement}

The part of this work due to D.~K.~has been supported
by a research grant within the
scope of the {\em Common Special Academic Program III} of the
Federal Republic of Germany and its Federal States, mediated 
by the DAAD.\\ 
The authors would like to thank L.~Vanzo for helpful discussions.

\appendix

\section{Vierbein and spin connection}\label{appvielbein}

In this section we give the choice of the vierbein and the spin connection
used in this paper for the charged rotating AdS black objects.
They solve the first Cartan equation
$\dd\ee^a+\omega^a{}_b\wedge\ee^b=0$. 
Setting $a=0$ in these equations, one recovers the vierbein and the spin
connection for the nonrotating black holes.

\subsection{Kerr-Newman-AdS Black Hole}\label{appKNAdS}

One choice of the vierbein for the Kerr-Newman-AdS black hole \rif{KNAdS}
is given by
\eq
  \ee^0={\sqrt{\Delta_r}\over\Xi\rho} (\dd t-a\sin^2\theta\dd\phi),\qquad
  \ee^1={\rho\over\sqrt{\Delta_r}}\dd r,
\feq
\eq
  \ee^2={\rho\over\sqrt{\Delta_\theta}}\dd\theta,\qquad
  \ee^3={\sqrt{\Delta_\theta}\sin\theta\over\Xi\rho} (a\dd t-(r^2+a^2)\dd\phi).
\feq
This implies the spin connection
\eq
\omega_t{}^{01}=\frac
1{2\Xi\rho^4}\left[\rho^2\Delta'_r-2r\Delta_r+2a^2r\sin^2\theta\Delta_\theta\right],
\feq
\eq
\omega_\phi{}^{01}=-\frac{a\sin^2\theta}{2\Xi\rho^4}
\left[\rho^2\Delta'_r-2r\Delta_r+2r(r^2+a^2)\Delta_\theta\right],
\feq
\eq
\omega_\phi{}^{02}=-{a\sqrt{\Delta_r\Delta_\theta}\over\Xi\rho^2}\sin\theta\cos\theta,
\feq
\eq
\omega_r{}^{03}={ar\sin\theta\over\rho^2}\sqrt{\Delta_\theta\over\Delta_r}\
,\qquad
\omega_\theta{}^{03}=-\frac{a\cos\theta}{\rho^2}\sqrt{\Delta_r\over\Delta_\theta},
\feq
\eq
\omega_r{}^{12}=-{a^2\over\rho^2}\sqrt{\Delta_\theta\over\Delta_r}\sin\theta\cos\theta,\qquad
\omega_\theta{}^{12}=-{r\over\rho^2}\sqrt{\Delta_r\over\Delta_\theta},
\feq
\eq
\omega_\phi{}^{13}={r\sqrt{\Delta_r\Delta_\theta}\over\Xi\rho^2}\sin\theta,
\feq
\eq
\omega_t{}^{23}=-\frac
a{2\Xi\rho^4}\left[\rho^2\Delta'_\theta\sin\theta+2(r^2+a^2)\Delta_\theta\cos\theta-2\Delta_r\cos\theta\right],
\feq
\eq
\omega_\phi{}^{23}=\frac
1{2\Xi\rho^4}\left[\rho^2(r^2+a^2)\Delta'_\theta\sin\theta+2(r^2+a^2)^2\Delta_\theta\cos\theta-2a^2\Delta_r\sin^2\theta\cos\theta\right].
\feq

\subsection{Rotating Generalization of the $g>1$ Charged
Topological Black Hole}\label{appHG}

One choice of the vierbein for the rotating generalization \rif{HG} of the
charged AdS black hole of genus
$g>1$ is given by
\eq
  \ee^0={\sqrt{\Delta_r}\over\Xi\rho} (\dd t+a\sinh^2\theta\dd\phi),\qquad
  \ee^1={\rho\over\sqrt{\Delta_r}}\dd r,
\feq
\eq
  \ee^2={\rho\over\sqrt{\Delta_\theta}}\dd\theta,\qquad
  \ee^3={\sqrt{\Delta_\theta}\sinh\theta\over\Xi\rho}(a\dd t-(r^2+a^2)\dd\phi),
\feq
leading to the spin connection
\eq
\omega_t{}^{01}=\frac
1{2\Xi\rho^4}\left[\rho^2\Delta'_r-2r\Delta_r+2a^2r\sinh^2\theta\Delta_\theta\right],
\feq
\eq
\omega_\phi{}^{01}=\frac{a\sinh^2\theta}{2\Xi\rho^4}
\left[\rho^2\Delta'_r-2r\Delta_r-2r(r^2+a^2)\Delta_\theta\right],
\feq
\eq
\omega_\phi{}^{02}={a\sqrt{\Delta_r\Delta_\theta}\over\Xi\rho^2}\sinh\theta\cosh\theta,
\feq
\eq
\omega_r{}^{03}={ar\sinh\theta\over\rho^2}\sqrt{\Delta_\theta\over\Delta_r}\
,\qquad
\omega_\theta{}^{03}={a\over\rho^2}\sqrt{\Delta_r\over\Delta_\theta}\cosh\theta,
\feq
\eq
\omega_r{}^{12}={a^2\over\rho^2}\sqrt{\Delta_\theta\over\Delta_r}\sinh\theta\cosh\theta,\qquad
\omega_\theta{}^{12}=-{r\over\rho^2}\sqrt{\Delta_r\over\Delta_\theta},
\feq
\eq
\omega_\phi{}^{13}={r\sqrt{\Delta_r\Delta_\theta}\over\Xi\rho^2}\sinh\theta,
\feq
\eq
\omega_t{}^{23}=-\frac
a{2\Xi\rho^4}\left[\rho^2\Delta'_\theta\sinh\theta+2(r^2+a^2)\Delta_\theta\cosh\theta+2\Delta_r\cosh\theta\right],
\feq
\eq
\omega_\phi{}^{23}=\frac
1{2\Xi\rho^4}\left[\rho^2(r^2+a^2)\Delta'_\theta\sinh\theta+2(r^2+a^2)^2\Delta_\theta\cosh\theta-2a^2\Delta_r\sinh^2\theta\cosh\theta\right].
\feq

\subsection{Charged Rotating Cylindrical Black Hole}\label{approttorus}

The vierbein used is this paper for the charged rotating cylindrical
black hole \rif{rottorus} is
\eq
  \ee^0={\sqrt{\Delta_r}\over\rho} (\dd t+aP^2\dd\phi),\qquad
  \ee^1={\rho\over\sqrt{\Delta_r}}\dd r,
\feq
\eq
  \ee^2={\rho\over\sqrt{\Delta_P}}\dd P,\qquad
  \ee^3={\sqrt{\Delta_P}\over\rho} (a\dd t-r^2\dd\phi),
\feq
which yields the spin connection
\eq
\omega_t{}^{01}=\frac
1{2\rho^4}\left[\rho^2\Delta'_r-2r\Delta_r+2ra^2\Delta_P\right],
\feq
\eq
\omega_\phi{}^{01}=\frac 1{2\rho^4}
\left[\rho^2aP^2\Delta'_r-2raP^2\Delta_r-2r^3a\Delta_P\right],
\feq
\eq
\omega_\phi{}^{02}={aP\sqrt{\Delta_r\Delta_P}\over\rho^2},
\feq
\eq
\omega_r{}^{03}={ar\over\rho^2}\sqrt{\Delta_P\over\Delta_r}\
,\qquad
\omega_P{}^{03}={aP\over\rho^2}\sqrt{\Delta_r\over\Delta_P},
\feq
\eq
\omega_r{}^{12}={a^2P\over\rho^2}\sqrt{\Delta_P\over\Delta_r},\qquad
\omega_P{}^{12}=-{r\over\rho^2}\sqrt{\Delta_r\over\Delta_P},
\feq
\eq
\omega_\phi{}^{13}={r\sqrt{\Delta_r\Delta_P}\over\rho^2},
\feq
\eq
\omega_t{}^{23}=-\frac
a{2\rho^4}\left[\rho^2\Delta'_P-2a^2P\Delta_P+2\Delta_rP\right],
\feq
\eq
\omega_\phi{}^{23}=\frac
1{2\rho^4}\left[\rho^2r^2\Delta'_P-2a^2r^2P\Delta_P-2a^2P^3\Delta_r\right].
\feq

\section{Real Representation of Dirac Matrices}\label{appgamma}

A real representation of Dirac matrices can be obtained by
applying a unitary transformation $U$ to the standard representation
(here denoted by $\gamma_a'$):
\eq
\ga_a = U^{-1}\ga_a' U,
\feq
where the transformation matrix $U$ is given by
\eq
U = \frac{1}{\sqrt{2}}\left(\begin{array}{cc}
                      1 & \sigma_2 \\
                      \sigma_2 & -1
                      \end{array} \right).
\feq
(The $\sigma_i$ denote the standard representation of the
two-dimensional Pauli matrices, i.~e.~$\sigma_3 = \mathrm{diag}(1,-1)$ etc.).
Using this, one obtains
\eqn 
\displaystyle
&\ga_0 = \left(\begin{array}{cc}
                      0 & -i\sigma_2 \\
                      -i\sigma_2 & 0
                      \end{array} \right),\qquad
&\ga_1 = \left(\begin{array}{cc}
                      -\sigma_3 & 0 \\
                      0 & -\sigma_3
                      \end{array} \right), \nonumber \\
&\ga_2 = \left(\begin{array}{cc}
                      0 & -i\sigma_2 \\
                      i\sigma_2 & 0
                      \end{array} \right), \qquad
&\ga_3 = \left(\begin{array}{cc}
                      \sigma_1 & 0 \\
                      0 & \sigma_1
                      \end{array} \right), \\
&\ga_5 = \ga_{0123} = \left(\begin{array}{cc}
                       i\sigma_2 & 0 \\
                       0 & -i\sigma_2
                       \end{array} \right). \nonumber
\feqn

\end{document}